\documentclass[12pt]{article}
\usepackage{amsmath, amssymb,amsbsy,amstext, amsthm, simplewick, amsfonts}
\usepackage{hyperref}
\usepackage{upgreek}
\usepackage{framed}
\usepackage{bbm}
\usepackage{textcomp}
\usepackage{adjustbox}
\usepackage{makecell}
\usepackage{physics}
\usepackage{empheq}
\usepackage[normalem]{ulem}
\usepackage{cite} % to get the citations with a dash [1-137]
\usepackage{verbatim}

\usepackage{braket}
\usepackage{tikz}
\usepackage{authblk}
\numberwithin{equation}{section}

\newcommand{\be}{\begin{eqnarray} }
\newcommand{\ee}{\end{eqnarray} }
\newcommand{\bs}{\begin{split} }
\newcommand{\es}{\end{split} }

\definecolor{color2}{rgb}{0.368417, 0.506779, 0.709798}
\definecolor{color3}{rgb}{0.880722, 0.611041, 0.142051}
\definecolor{color5}{rgb}{0.560181, 0.691569, 0.194885}
\definecolor{color1}{rgb}{0.922526, 0.385626, 0.209179}
\definecolor{color6}{rgb}{0.528488, 0.470624, 0.701351}
\definecolor{color4}{rgb}{0.772079, 0.431554, 0.102387}

\begin{document}

\begin{titlepage}
\setcounter{page}{1} \baselineskip=15.5pt 
\thispagestyle{empty}

\begin{center}
{\fontsize%{28}{28} 
{28}{28}\centering \bf Stringy Completions of the  \\ \vspace{0.3cm}  Standard Model from the Bottom Up}\\
\end{center}

\vskip 38pt
\begin{center}
\noindent
{\fontsize{12}{18}\selectfont Brad Bachu\footnote{\tt bbachu@princeton.edu} and Aaron Hillman\footnote{\tt aaronjh@princeton.edu}}
\end{center}

\begin{center}
\vskip 8pt
\textit{Department of Physics, Princeton University, Washington Road, Princeton, NJ, USA}
\end{center}

%=========================================

\vspace{2.4cm}

\noindent We study a class of tree-level ans{\"a}tze for $2\to 2$ scalar and gauge boson amplitudes inspired by stringy UV completions.  These amplitudes manifest Regge boundedness and are exponentially soft for fixed-angle high energy scattering, but unitarity in the form of positive expandability of massive residues is a nontrivial consistency condition. In particular, unitarity forces these ans{\"a}tze to include graviton exchange.  In the context of gauge boson scattering, we study gauge groups $SO(N)$ and $SU(N)$. In four dimensions, the bound on the rank of the gauge group is $24$ for both groups, and occurs at the maximum value of the gauge coupling $g_{YM}^2 = \frac{2M_s^2}{ M_P^2} $.  In integer dimensions $ 5\leq   D \leq 10$ , we find evidence that the maximum allowed allowed rank $r$ of the gauge group agrees with the swampland conjecture $r < 26-D$.  The bound is surprisingly identical for both $SU(N)$ and $SO(N)$ in integer spacetime dimensions.  We also study the electroweak sector of the standard model via $2 \to 2$ Higgs scattering and find interesting constraints relating standard model couplings, the putative string scale, and the Planck scale.

%=========================================

\end{titlepage} 

%\restoregeometry

\newpage
\setcounter{tocdepth}{2}
{
\hypersetup{linkcolor=black}
\tableofcontents
}

\newpage

%=======================================
%======================================

%%%%%%%%%%%%%%%%%%%%%%%%%%%%%%%%%%%%%

\section{Introduction} 
Already at tree level one can diagnose the need for the UV completion of amplitudes with graviton exchange.  Such amplitudes grow with center-of-mass energy and need to be unitarized below the Planck scale.  This problem is of course not unique to gravitational amplitudes; scattering of longitudinally polarized $W$ bosons exhibits the same growth with energy, a violation of unitarity which we now understand to be remedied by a weakly coupled Higgs.  What is more surprising is that the force we have known the longest has proven the hardest to UV complete.\\  This state of affairs can be understood in the context of unitarity constraints on massless scattering amplitudes. From this point of view, the graviton mediating a universally attractive force is understood as a consequence of consistent factorization of massless scattering amplitudes with a helicity two particle. By dimensional analysis, the helicity demands gravity has an irrelevant coupling and an associated growth in energy for amplitudes exchanging gravitons.  So, consistency of massless scattering both make sense of gravity's privileged position as our oldest force on the books and implies it is the most immediately problematic long-range force at high energies.\\  But why is it harder to UV complete than $W$ scattering?  \textit{Because} it is a long-range force.  High energy growth of the amplitude is not the only way to violate unitarity; it can be violated at low energies as well, by not having a positively expandable imaginary part.  Contact interactions produce neither singularities nor imaginary parts for tree level amplitudes and therefore do not directly impose additional \textit{a priori} constraints on the amplitude.  This affords a certain freedom in engineering UV-completions by resolving a contact interaction into particle production.  This is not the case for amplitudes with massless exchanges.  The same consistency conditions which imply gravity's universal attraction are positivity constraints which must be respected by any putative UV-completion.  This proves to be a surprisingly stringent constraint.\\  But we also know gravity need not merely be a fly in our unitarity ointment; in the context of string theory, gravity is $\textit{required}$ for unitarity. And by studying consistency conditions on amplitudes from the right point of view, we can see in an analogous way that gravity is not an obstruction, but rather a facilitator of certain means of unitarization.  That is what we find in particular for the class of ans{\"a}tze considered in this work. And by considering stringy ans{\"a}tze motivated by amplitudes with graviton exchange, we can initiate a program of building UV completions of the standard model (SM) from the bottom up.   \\
The outline of this work is as follows.  In Sec. \ref{sec:unit} we review unitarity constraints on amplitudes and motivate a certain class of stringy ans{\"a}tze for $2\to 2$ scattering which unitarize field theory amplitudes.  These amplitudes resemble closed string amplitudes, but are strictly studied from the point of view of unitarity constraints on $2 \to 2$ scattering, not any explicit realization in some string background.   In Sec. \ref{sec:gaugeboson}, with the goal of studying standard model and beyond the standard model (BSM) amplitudes, we study $SO(N)$ and $SU(N)$ gauge boson amplitudes.  While in this neighborhood, we also probe unitarity constraints in general spacetime dimensions, finding evidence for example that the maximum allowed $N$ for $SO(N)$ is 32 as realized by the $SO(32)$ heterotic string. Finally, in Sec. \ref{sec:SM} we study Higgs scattering and comment on the compatibility of our amplitudes with the SM. \\

\section{Unitarity and UV Completion}
\label{sec:unit}
Before moving on to specific ans{\"a}tze, it is useful to review the constraints of Lorentz invariance and unitarity on the amplitude and some well-known instances of tree-level UV-completion in the absence of gravity.  This will illustrate the fundamental challenge with even tree-level UV-completion in the presence of gravity and motivate the stringy form factor which will dress all the amplitudes considered herein.  We always consider amplitudes with massless external states and employ spinor helicity with mandelstams
\begin{equation}
	s = 2p_1\cdot p_2 \hspace{14mm} t = 2p_2\cdot p_3 \hspace{14mm} u = 2p_1 \cdot p_3
\end{equation}
and $s+t+u = 0$.  For on-shell kinematics the spinor-brackets obey $\braket{ij} = \pm [ij]^\star$ with the sign depending whether the states are ingoing or outgoing.  When imposing positive expandability on a basis of orthogonal polynomials, the argument of the polynomials is $\cos\theta$ where 
\begin{equation}
	t = -\frac{s}{2}(1-\cos\theta) \hspace{15mm} u = -\frac{s}{2}(1+\cos\theta)
\end{equation}   
\subsection{Review of Unitarity Constraints}
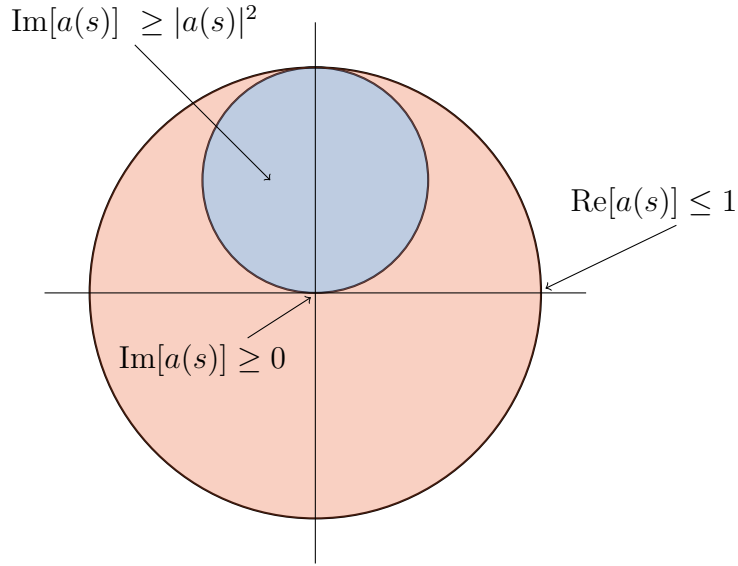
\begin{figure}[h!]
	\centering
	\begin{tikzpicture}[scale=.3]
		\draw[thick] (0, 5) circle (50mm);
		\draw[fill, color = color2, opacity=.4] (0, 5) circle (50mm);
		\draw[thick] (0, 0) circle (100mm);
		\draw[fill, color = color1, even odd rule, opacity=.3] (0, 0) circle (100mm) (0, 5) circle (50mm);
		\draw[] (0, -12) -- (0, 12);
		\draw[] (-12, 0) -- (12, 0);
		\node at (-5, -3) {$\text{Im}[a(s)] \geq 0$};
		\draw[->] (-3, -2) -- (-.2, -.2);
		\node at (15, 4) {$\text{Re}[a(s)] \leq 1$};
		\draw[->] (16, 3) -- (10.2, .2);
		\node at (-8, 12) {$\text{Im}[a(s)]\text{  } \geq |a(s)|^2$};
		\draw[->] (-8, 11) -- (-2, 5);
	\end{tikzpicture} 
	\caption{Above we depict in the complex $a(s)$ the bounds relevant for our analysis.  The non-perturbative bound on $a(s)$ is shaded in blue.  At weak coupling, one can diagnose unitarity violation at large $s$ using the weaker, pink shaded condition.  We then impose positivity of the imaginary part which is the weak-coupling portion of the blue shading (where $|a(s)|^2$ is parametrically suppressed. )}
	\label{fig:argandcircle}
\end{figure}
 Lorentz invariance allows us to expand the elastic two-to-two amplitude for massless particles as
\begin{equation}
	\mathcal{A}(1_{r_1}^{h_1}, 2_{r_2}^{h_2}, 3_{r_3}^{h_3}, 4_{r_4}^{h_4}) = 16\pi\sum\limits_J (2J+1)a_{J, R}^{\{h_i\}}(s) \mathbb{G}^{\{ h_i\}}_{D,J}(\cos\theta)\mathbb{P}_R(\{ r_i \})
	\label{eq:expansion}
\end{equation}
The $\mathbb{G}^{\{ h_i\}}_{D,J}$ are the relevant basis of orthogonal (in general spinning) polynomials for the scattering process in question, corresponding to spin-$J$ exchange in $D$ space-time dimensions with external helicities $\{h_i\}$.  We will state the particular polynomial basis for each process we consider.  The $\mathbb{P}_R(\{ r_i \})$ are projectors in the internal symmetry space for representations $r_i$ exchanging representation $R$ (in the $s$-channel here).  The weights $a_{J, R}^{\{h_i\}}(s)$ in front of these two sets of orthogonal projectors are partial wave coefficients, which depend on Mandelstam $s$ and the constants in the amplitude. Unitarity then imposes the bound on partial waves
\begin{equation}
	\label{eq:unitaritybound}
	\text{Im}[a(s)] \geq |a(s)|^2
\end{equation}
where the constraint holds for each $a_{J, R}^{\{h_i\}}(s)$ individually and we suppress the labels on $a(s)$.\footnote{The $S$-matrix is required to be a positive operator in general, therefore in general this constraint is a statement about positivity of a matrix, but we will be studying amplitudes for which each individual $a(s)$ obeys (\ref{eq:unitaritybound}).}  Noting that this equation is identical to 
\begin{equation}
	\left(\text{Im}[a(s)]-1/2
	\right)^2+\text{Re}[a(s)]^2 \leq \frac{1}{4}
\end{equation}
the unitarity bound clearly forces each $a(s)$ to lie in the Argand circle depicted in figure \ref{fig:argandcircle}.  Two limits of this non-perturbative constraint will be useful for our weak coupling analysis, each simplifying either the left or right-hand side of equation (\ref{eq:unitaritybound}).  These limits are
\begin{equation}
	\begin{cases}
		\text{Im}[a(s)]  \geq 0  & \text{ positive expandability at singular loci}\\
		|a(s)| \leq 1 & \text{ boundedness at large $s$}
	\end{cases}
\end{equation}
The tension between these two conditions and causality imposed via Regge boundedness in the complex $s$ plane, places stringent constraints on the form of weakly-coupled gravitational UV completions. We will be considering processes with graviton exchange, which exhibit unitarity violation at large $s$ and need to be UV completed below the Planck scale. Rather than give up on weak coupling, we will posit an ansatz for a weakly coupled completion which softens the high-energy behavior.  This will be done by the introduction of new heavy states and positive expandability at the associated singularities must be checked.  This constrains the low-energy data in the context of the specific ansatz to which we have committed.  In this work we will manifest all necessary conditions on the amplitude other than positive expandability of residues, leaving it as the non-trivial condition which must be checked.\\ 
In addition to boundedness at fixed angle, positive expandibility of the imaginary part, and Regge boundedness, we impose additional reasonable constraints on the spectrum.  In particular we require that there are a fixed number of spins at a given mass i.e. that the Chew-Frautschi plot has no spikes.  We summarize the full set of constraints below
\begin{enumerate}
	\item[$\circ $] Boundedness at fixed angle and high energy via
	\begin{equation}
		|a(s)| \leq 1
	\end{equation}
	\item[$\circ $] Regge boundedness for complex $s$:
	\begin{equation}
		\lim_{|s|\to \infty} \mathcal{A}/s^2 \to 0 \hspace{15mm} \text{ fixed } t < 0
	\end{equation}
	\item[$\circ $] Fixed number of spins at a given mass by imposing that any residue in $s$ is a polynomial in $t$.
	\item[$\circ $] Positive expandability of the imaginary part, the only condition which we will not manifest.
\end{enumerate} 
\subsection{Completions}
Before moving on to constructing gravitational completions, it is useful to recapitulate a well-known example of a non-gravitational tree level amplitude requiring UV completion: the non-linear sigma model.  This will help frame the unique challenge and opportunities afforded in studying gravitational completions.
\paragraph{Non-linear Sigma Model:} We consider the amplitude for four Goldstones
\begin{multline}
	\label{eq:ampnlsm}
	\mathcal{A}_{\text{NLSM}}  = -\frac{1}{f^2}\bigg[(s+t)\left(\text{Tr}(T^a T^b T^c T^d)+\text{Tr}(T^d T^c T^b T^a)\right)+(s+u)(\text{Tr}(T^b T^a T^d T^c)+\text{Tr}(T^c T^d T^a T^b))\\+(t+u)(\text{Tr}(T^a T^c T^b T^d) +\text{Tr}(T^d T^b T^c T^a) )\bigg]
\end{multline}
With the $T^a$ standard generators of the defining representation of $SU(N)$.  We have scalar partial wave for any exchanged representation
\begin{equation}
	a_0(s) \sim \frac{s}{f^2}
\end{equation}
which means the amplitude needs to be unitarized before $s \sim f^2$.  Already in this case we can illustrate how the tensions between boundedness at fixed-angle and Regge boundedness in the complex plane don't allow the most naive softenings of the amplitude.  For example, we could try the exponential form factor
\begin{equation}
	\label{eq:brutal}
	\hat{\mathcal{A}} = e^{-\frac{s^2+t^2+u^2}{M^4}}\mathcal{A}_{\text{NLSM}}
\end{equation}
which is exponentially soft at fixed angle, but diverges exponentially for imaginary $s$ at fixed $t$. If we settle on more humble expectations for softening, the most obvious way to soften the amplitude is to divide it by something i.e. introduce a massive exchange, which should in particular have positive mass-squared.  Therefore the simplest ansatz is
\begin{equation}
	\label{eq:ampnlsmUV}
	\hat{ \mathcal{A}}_{\text{NLSM}}  = -\frac{1}{f^2}\left[\frac{s}{1-\frac{s}{M^2}} \left(\mathbb{P}_{1}^s+ \mathbb{P}_{2}^s\right)+\frac{t}{1-\frac{t}{M^2}} \left(\mathbb{P}_{1}^t+ \mathbb{P}_{2}^t\right)+\frac{u}{1-\frac{u}{M^2}} \left(\mathbb{P}_{1}^u+ \mathbb{P}_{2}^u\right)\right]
\end{equation}
where we decomposed the usual flavor-ordered form into the form manifesting exchanges in respective channels, with
\begin{equation}
	\label{eq:pioncolor}
	 \begin{cases}
	 	\mathbb{P}_1^s  = \frac{2}{N} \delta_{ab}\delta_{cd}\\
	 \mathbb{P}_2^s  =d_{abe}d_{cde}
	 \end{cases}
\end{equation}
which correspond to the singlet and symmetric adjoint irreducible flavor exchanges.  Now, the partial waves are clearly bounded so long as $M^2 < f^2$ i.e. the UV completion scale is below the UV cutoff $f^2$.  But we are not done, the amplitude now has a singularity at $M^2$ in each channel and the residue must be positive.  This additionally gives the condition $f^2 > 0$.  The low-energy amplitude alone was agnostic about the sign of $f^2$, but once committed to a certain UV completion, it is possible to constrain low-energy data by imposing positive expandability of residues.  We will find in the case of gravitational amplitudes that constructing a UV completion is more difficult, the upshot being that the constraints on low energy data are far more interesting.
\paragraph{Graviton Exchange:} Our objective is to study UV completions in the presence of graviton exchange, so let's see if we can apply the lessons from the non-linear sigma model to the case of four identical scalars exchanging gravitons.  The amplitude is 
\begin{equation}
	\label{eq:fourscalar}
	\mathcal{A}_{\text{grav}} = -\frac{1}{M_P^2} \left(\frac{t u}{s}+\frac{s u}{t}+\frac{s t}{u} \right)
\end{equation}
Where $M_P$ is the reduced Planck mass $M_P^{-2} = 8\pi G $. Again we see that the scalar partial wave grows with energy
\begin{equation}
	a_0(s) \sim \frac{s}{M_P^2}
\end{equation}
As with the NSLM amplitude, we do not want to give up on perturbativity; we will pursue a tree level UV completion.  It was easy to engineer a unitarization of the NLSM amplitude, what could be the difficulty with gravity?  Particle production at low energies.  More explicitly, unlike the NLSM amplitude, the graviton exchanges already impose a sign condition on the coupling squared.  This fixed sign in the low-energy amplitude is something our putative completion will have to respect, and it has surprisingly drastic consequences.  The issue has everything to do with these low energy poles, and can already be seen in the context of $\phi^3$ amplitudes. Suppose we wanted to UV-improve the behavior of the $\phi^3$ amplitude; we can try to mimic the strategy employed in the NLSM.  Focusing on a single channel
\begin{equation}
	\label{eq:phi3uvimprove}
	\frac{g^2}{s} \to \frac{g^2}{s} \times \frac{1}{\prod\limits_{i = 1}^{n}\left(1-\frac{s}{M_{s, i}^2}\right) }
\end{equation}
where we have introduced $n$ new massive poles to soften the high-energy behavior to $\frac{1}{s^{n+1}}$.  As with the initial $s = 0$ pole, we require the residue on each of these massive poles to be positive.  But notice that because there is no pole at infinity, a residue theorem guarantees that some of the residues must have the wrong sign.  We have arrived at the need for an infinite number of poles to have a hope of softening this amplitude in a way consistent with unitarity.  We can make an ansatz for a factor with an infinite number of poles in each channel which will dress the amplitude:
\begin{equation}
	\mathcal{D}(s, t, u) = \frac{N(s, t, u)}{\prod\limits_i \left(s-M_{s, i}^2 \right)\prod\limits_j \left(t-M_{t, j}^2 \right)\prod\limits_k \left(u-M_{u, k}^2 \right)}
\end{equation}
In order to further constraint this putative dressing factor, we will impose the well-motivated condition that there is a finite number of spins at given mass-level.  Analytically, this means that the residue at each massive pole must be a polynomial in the remaining Mandelstam invariant, in particular the numerator must cancel the poles in the other channel upon taking such a residue.  The denominator is a product of polynomials so it's natural to have the same ansatz for the numerator 
\begin{equation}
	\label{eq:numerator}
	N(s, t, u) =  \prod\limits_i(s+r_{s,i})\prod\limits_j(t+r_{t,j})\prod\limits_k(u+r_{u,k})
\end{equation}
The residue is then 
\begin{equation}
	\label{eq:gammres}
	\text{Res}_{s \to M_{s, i}^2}\mathcal{D} \propto \frac{ \prod\limits_j(t+r_{t,j})\prod\limits_k\left(t+M_{s, i}^2-r_{u,k} \right)}{ \prod\limits_j \left(t-M_{t, j}^2 \right)\prod\limits_k \left(t+M_{u, k}^2+M_{s,i}^2 \right) }
\end{equation}
requiring cancellation of the remaining poles in all channels yields the condition:
\begin{equation}
	\label{eq:mrcondition}
	M_{s, i}^2 +M_{t,j}^2 \in \{ r_{u, k} \}
\end{equation}
plus its cyclic rotations.  We will proceed with the most obvious way of solving this constraint, which is to make all six of these putative sets one mass scale times the positive integers.  We will call this mass scale $M_s$ in which case we find 
\begin{equation}
	\label{eq:dresssolve}
	\mathcal{D}(s, t, u) = \frac{ \prod\limits_{i =1}^\infty (s+M_s^2 i)\prod\limits_{j =1}^\infty (t+M_s^2  j)\prod\limits_{k =1}^\infty (u+M_s^2 k)}{\prod\limits_{i =1}^\infty (s-M_s^2 i)\prod\limits_{j =1}^\infty (t-M_s^2  j)\prod\limits_{k =1}^\infty (u-M_s^2 k)}
\end{equation}
Which we recognize as a famous function
\begin{equation}
	\label{eq:gammastr}
	\Gamma^{\text{str}} = -\frac{\Gamma\left(-\alpha' s \right)\Gamma\left(-\alpha' t \right)\Gamma\left(-\alpha' u \right)}{\Gamma\left(\alpha' s \right)\Gamma\left(\alpha' t \right)\Gamma\left(\alpha' u \right)}
\end{equation}
with $\alpha' = \frac{1}{M_s^2}$.  Though we did not impose these conditions, it is readily verified from the form in (\ref{eq:gammastr}) that the amplitude satisfies Regge boundedness and boundedness at fixed angle and high energy. We summarize the properties of this amplitude below:
\begin{enumerate}
	\item[$\circ $] Exponentially soft at high-energy, fixed angle
	\item[$\circ $] Regge limit:
	\begin{equation}
		\lim_{|s|\to \infty} \Gamma^{\text{str}} = s^{\alpha' t} \hspace{15mm} \text{ fixed } t < 0
	\end{equation} 
	\item[$\circ$] A residue in $s$ at level $n$ i.e. mass $M^2 = n M_s^2$ is a polynomial of degree $2n$ in $t$ and is in particular 
	\begin{equation}
	\label{eq:gammares}
		\text{Res}_{s\to n/\alpha'}  \Gamma^{\text{str}} = \frac{1}{n!(n-1)!}\left(\prod\limits_{i = 1}^{n-1}(i+\alpha' t)^2\right)t(n+\alpha' t)
	\end{equation} 
\end{enumerate}
 If we dress our graviton exchange amplitude (\ref{eq:fourscalar}) with $\Gamma^{\text{str}}$ this is in fact the four dilaton amplitude in type IIB.  This quasi-derivation of the Virasoro-Shapiro amplitude is far from new; Virasoro remarked on it in his original paper after all \cite{Virasoro:1969me}, but it is surprising.  Though we by no means claim to be proving that this is the unique way of unitarizing graviton exchange, it is remarkable that with very little input and at each stage making the simplest of moves, we arrive at a form factor with Regge boundedness and \textit{exponential} softness at fixed angle. These two conditions are extremely non-trivial to engineer, but we were able to build an amplitude with both properties by seeking a UV completion in the presence of gravity.  We will find that the arrow goes both ways.  Gravity does not merely necessitate the discovery of this somewhat sophisticated completion, it is in fact necessary for the use of such a completion.
\paragraph{Partial Wave Unitarity:}With an ansatz in hand for gravitational amplitudes, we can now impose positive expandability of the residues.  The amplitude is by construction positive on the massless poles, so we only need to check the positive expandability on the new poles in the completion.  These come from $\Gamma^\text{str}$ and the residue of $\Gamma^\text{str}$ at general mass level $n$ was stated above in (\ref{eq:gammares}).\\
\indent Now that we have $\Gamma^{\text{str}}$ and all of its wonderful properties at our disposal, one might hope that with this unitarizing hammer, any field theory amplitude looks like a nail.  We can test it on something simple such as $\phi^4$ theory, in which case we merely check the partial wave expansion of $\Gamma^{\text{str}}$.  At the first mass-level one can readily verify that 
\begin{equation}
	\text{Res}_{s\to 1/\alpha'} \Gamma^{\text{str}} \propto \bigg(P_2(\cos\theta)\color{red}-P_0(\cos\theta)\color{black} \bigg)
\end{equation}
so we find a negative residue.  We could try coupling it to gravity; the amplitude is just augmenting the four-dilaton amplitude in type IIB by a contact interaction:
\begin{equation}
	\label{eq:fourdillambda}
	\mathcal{A}^\lambda = \Gamma^{\text{str}}\left[-\frac{1}{M_P^2}  \left(\frac{t u}{s}+\frac{s u}{t} +\frac{s t}{u}\right)+\lambda\right]
\end{equation}
We have
\begin{equation}
	\text{Res}_{s\to 1/\alpha'} \mathcal{A}^\lambda \propto \frac{1}{70}P_4(\cos\theta)+\left(\frac{2}{7}+\frac{\alpha'M_P^2\lambda}{6} \right)P_2(\cos\theta)+\left(\frac{7}{10}-\frac{\alpha'M_P^2\lambda}{6} \right)P_0(\cos\theta)
\end{equation}
 And so we find the two-sided bound on $\lambda$:
 \begin{equation}
 \frac{-12}{7}	\leq \lambda \frac{M_P^2}{M_s^2} \leq \frac{21}{5}
 \end{equation}
So we can have a quartic coupling so long as it does not overwhelm the graviton exchange piece of the amplitude.  In this sense, we are able to UV-improve $\lambda \phi^4$ theory at high energies so long as we couple it to gravity. In the context of these ans{\"a}tze, gravity is not an obstruction to unitary UV softening, but a necessity.  One could imagine that perhaps this was merely the absence of long-distance physics at low energies that was the problem, not gravity in particular.  We can test this on pure gauge boson scattering.  Checking unitarity in this case requires an additional layer of calculation: 6-$j$ coefficients.
\paragraph{6-$j$ Coefficients:}  Since we are working at tree level, taking the imaginary part merely amounts to taking the residue (\ref{eq:gammres}).  Moreover, in practice we find that the strongest constraints come at low mass levels, stabilizing at the latest by mass level three for the amplitudes considered in this analysis.  The most non-trivial aspect of testing positive expandability is in re-expanding the color structures of (\ref{eq:fourgaugeboson}), which as such are not expressed in the basis of $s$-channel projectors.  The full set of projectors in any one of the three channels, full set meaning the projectors for all representations that can be exchanged in this channel, are an orthogonal basis of projectors depending on the external indices.  Therefore, each projector in the $t$ and $u$ channels can be uniquely expanded in terms of the projectors in the $s$-channel:
\begin{equation}
\label{eq:pexpand}
	\mathbb{P}_R^{t} = \sum\limits_{R'} C^{t, s}_{R, R'} \mathbb{P}_{R'}^s
\end{equation} 
where we have suppressed the external indices which these projectors are functions of.  This is merely the crossing the equation, and it is solved trivially by contraction given that the basis of projectors corresponding to the exchange of irreps $R$ are orthogonal.  That is
\begin{equation}
	\label{eq:psolve}
	C^{t, s}_{R, R'}  = \frac{\mathbb{P}_R^{t}\cdot \mathbb{P}_{R'}^s}{ \mathbb{P}_{R'}^s \cdot  \mathbb{P}_{R'}^s}
\end{equation}
where the dot denotes contraction with the relevant invariants on the external indices which have been suppressed.  This coefficient $C^{t, s}_{R, R'}$ is a 6-$j$ symbol, depending on the four external representations in addition to the two exchanged representations $R$ and $R$'.  These must be calculated for the group and representations in question.  Once this is done, the projection onto irreducible exchanged states is straightforward.  A useful resource for the determination of these projectors is \cite{Cvitanovic:2008zz}. \\
We can return to analyzing our four gauge boson amplitude
\begin{equation}
\label{eq:puregaugeboson}
	\mathcal{A} = \frac{g_{YM}^2}{3} \braket{12}^2[34]^2\Gamma^{\text{str}}\left(\frac{\mathbb{P}_{\text{Adj}}^s-\mathbb{P}_{\text{Adj}}^t}{st}+\frac{\mathbb{P}_{\text{Adj}}^t-\mathbb{P}_{\text{Adj}}^u}{tu}+\frac{\mathbb{P}_{\text{Adj}}^u-\mathbb{P}_{\text{Adj}}^s}{su}\right)
\end{equation}
where we have the adjoint projectors 
\begin{equation}
\label{eq:adjproj}
	\mathbb{P}_{\text{Adj}}^s = f^{abe}f^{edc}
\end{equation}
and similarly for the $t$ and $u$ channels.  We can focus on the explicit case of $SO(N)$.   Positive expandability is already violated at level one for the most subleading Regge trajectory in the largest representation exchanged between the adjoints.  The exchanged state and associated coefficient in the partial wave expansion are:
\begin{equation}
\label{eq:puregaugeviol}
	\begin{tikzpicture}
	\node at (0, 0) {\includegraphics[scale=.6]{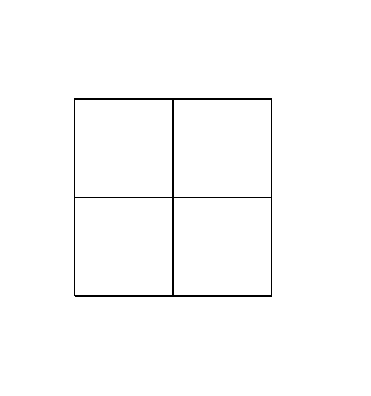} };
	\node at (1.2, -1 ) {\tiny $m^2 = 1, J = 0$};
	\node at (3, 0) {$:$};
	\node at (5, 0) {$-g_{YM}^2$};
\end{tikzpicture}
\end{equation}
 Where the left is the Young tableaux for the exchanged representation of $SO(N)$, the mass-squared in units of $M_s^2$, and the spin.  The coefficient is identical for the analogous representation of $SU(N)$.  Perhaps the most salient feature of this violation is the indication that the actual heterotic string amplitude becomes unitarity violating once gravity becomes \textit{too weak} relative to the gauge interactions; we cannot have the gauge interactions all on their own. Yet again we find that unitarity of these amplitudes requires graviton exchange.   String theory of course necessitates gravity, so unitarity violation in the complete absence of gravity, is perhaps not so surprising from that point of view.  But what is interesting about this analysis is that we can move around in the space of parameters untethered to any choice of background or class of compactifications and make what appear to be robust observations about constraints on this class of amplitudes with graviton exchange.  For instance, extending the above to our full ans{\"a}tze will reveal that the generalization of (\ref{eq:puregaugeviol}) is in tension, but compatible, with the weak gravity conjecture (WGC). 
\paragraph{Summary of Gravitational Ans{\"a}tze:} Now that we have motivated both the prefactor $\Gamma^{\text{str}}$ and that it must be used for completions along with graviton exchange, we can discuss more completely the rules bounding our ans{\"a}tze.  The basic game is clear: multiply a massless field theory amplitude by the stringy form-factor $\Gamma^{\text{str}}$ and check positive expandability.  Now we discuss the restrictions we impose on the massless field theory amplitudes.\\   In motivating $\Gamma^{\text{str}}$ we already took as assumptions that the residues of the amplitude in $s$ were polynomials in $t$.  Additionally, we impose Regge boundedness of the amplitude, which in the case of our gravitational amplitudes means bounded by $s^2$.  Seeing as $\Gamma^{\text{str}}$ goes as $s^{\alpha' t}$ in the Regge limit, the field theory amplitude which this factor multiplies can scale at most as $s^2$ in the Regge limit, which the graviton exchange piece indeed will.  This bounds the dimension of interactions coming from operators that we may put in by hand.  As for denominators, one can consider the heterotic-type deformation
\begin{align}
\label{eq:hetdef}
	\frac{1}{s} \to \frac{1}{s(1+\alpha' s)}
\end{align}  
It is crucial that this potential tachyon pole has vanishing residue
\begin{equation}
	\label{eq:hetpoleres}
	\text{Res}_{s\to -1/\alpha'} \frac{1}{s(1+\alpha' s)}\Gamma^{\text{str}} = 0
\end{equation}
Indeed any product of $(n+\alpha' s)$ for distinct integers $n$ can be added to the denominator and will obey the vanishing residue condition above.  But they will violate our condition for a finite number of spins at a given mass level.  Recall from (\ref{eq:gammres}) that at the first mass level we have 
\begin{equation}
	\label{eq:firstmassres}
	\text{Res}_{s\to 1/\alpha'}\Gamma^{\text{str}} \propto t(1+\alpha't) = u(1+\alpha'u)
\end{equation}
which means, considering the $t$ and $u$ channel terms in our amplitude, that we can have our massless poles and the $(1+\alpha't)$ or $(1+\alpha' u)$ factors as well, but no more.  Additional factors would fail to cancel upon taking the residue at the first mass level and introduce an infinite number of spins.  For simplicity, we only consider these poles on the graviton exchange part of the amplitude.\\  Therefore, we can compute a massless scattering amplitude, the form of which is itself fixed by consistent factorization into three-particle amplitudes on the massless poles.  This amplitude can have contact terms put in by hand up to scaling as $s^2$ in the Regge limit.  The amplitude is then dressed by $\Gamma^{\text{str}}$ which ensures that the amplitude satisfies all necessary criteria, except potentially positive expandability on the new massive poles.  Checking this condition produces constraints on the low energy data.
\section{Gauge Boson Scattering}
\label{sec:gaugeboson}
First we take up massless gauge boson scattering.  At tree-level, the non-trivial $2 \to 2$ gauge boson amplitudes are those of $W$'s and gluons.  There is a $2\to 2$ amplitude for $B$'s (gauge boson of $U(1)_Y$), but it is purely mediated by graviton exchange and therefore not further constrained by unitarity.  In addition to these, in many GUT models we have gauge bosons of $SO(N)$ or $SU(N)$, and we will carry out the analysis for general $N$ in both cases.  The massless scattering amplitude is
\begin{multline}
	\label{eq:fourgaugeboson}
	\mathcal{A}(1^{-a}, 2^{-b}, 3^{+c}, 4^{+d}) = \braket{12}^2[34]^2\bigg[\frac{1}{M_P^2}\left(\frac{\mathbb{P}_{1}^s}{s}+\frac{\mathbb{P}_{1}^t}{t}+\frac{\mathbb{P}_{1}^u}{u} \right)+ \\\frac{g_{\text{YM}}^2}{3} \left(\frac{\mathbb{P}_{\text{Adj}}^s-\mathbb{P}_{\text{Adj}}^t}{st}+\frac{\mathbb{P}_{\text{Adj}}^t-\mathbb{P}_{\text{Adj}}^u}{tu}+\frac{\mathbb{P}_{\text{Adj}}^u-\mathbb{P}_{\text{Adj}}^s}{su}\right)\bigg]
\end{multline}
we write the color structures in this way as they are the color structures to be expanded in on a factorization channel.  In particular, these are
\begin{equation}
	\begin{cases}
		\mathbb{P}_1^s  & \delta^{ab}\delta^{cd}\\
		\mathbb{P}_{\text{Adj}}^s & f^{abe}f^{edc}
	\end{cases}
\end{equation}
  Note that in the Regge limit, this amplitude already scales as $s^2$.  Any additional contribution consistent with Regge behavior would require new poles, which we do not have at low energies. Our ansatz is then 
\begin{equation}
	\mathcal{A}^{\text{UV}}(1^{-a}, 2^{-b}, 3^{+c}, 4^{+d}) = \Gamma^{\text{str}}\mathcal{A}(1^{-a}, 2^{-b}, 3^{+c}, 4^{+d})
\end{equation}
or $\mathcal{A}^{\text{UV}}_{\text{Het}}$ which deforms the graviton poles to look heterotic
\begin{multline}
	\label{eq:hetfourgaugeboson}
	\mathcal{A}^{\text{UV}}_{\text{Het}}(1^{-a}, 2^{-b}, 3^{+c}, 4^{+d}) = \braket{12}^2[34]^2\bigg[\frac{1}{M_P^2}\left(\frac{\mathbb{P}_{1}^s}{s(1+\alpha' s)}+\frac{\mathbb{P}_{1}^t}{t(1+\alpha't)}+\frac{\mathbb{P}_{1}^u}{u(1+\alpha'u)} \right)+ \\\frac{g_{\text{YM}}^2}{3} \left(\frac{\mathbb{P}_{\text{Adj}}^s-\mathbb{P}_{\text{Adj}}^t}{st}+\frac{\mathbb{P}_{\text{Adj}}^t-\mathbb{P}_{\text{Adj}}^u}{tu}+\frac{\mathbb{P}_{\text{Adj}}^u-\mathbb{P}_{\text{Adj}}^s}{su}\right)\bigg]
\end{multline}
We introduce $g_s$ via
\begin{equation}
\label{eq:gstring}
	g_s^2 = \frac{M_s^2}{M_P^2}
\end{equation}
which is the dimensionless strength characterizing how far below the Planck scale the UV completion scale $M_s$ is i.e. how weakly coupled a completion of gravity it is.
The strongest constraints come from the pole in the $s$-channel with the helicity configuration in (\ref{eq:fourgaugeboson}).  Furthermore, we note that this ansatz is perfectly good in any number of spacetime dimensions, and we will also consider the necessary but insufficient condition of positive expandability on the \textit{scalar} $D$-dimensional Gegenbauer polynomials.

\subsection{Constraints in Four Spacetime Dimensions}
In four dimensions, the polynomials are the spinning Gegenbauer's
\begin{equation}
	\label{eq:spinninggegs}
	\mathbb{G}_{D = 4, J}^{\{h_i\}}(\cos\theta) = d^J_{h_{12}, h_{34}}(\cos\theta)
\end{equation}
where $h_{ij} = h_i-h_j$.  In the $s$-channel for (\ref{eq:fourgaugeboson}) these polynomials actually correspond to the scalar Legendre's, and we find the strongest constraints.
\begin{figure}
 \hspace{-1cm}\includegraphics[scale=.365]{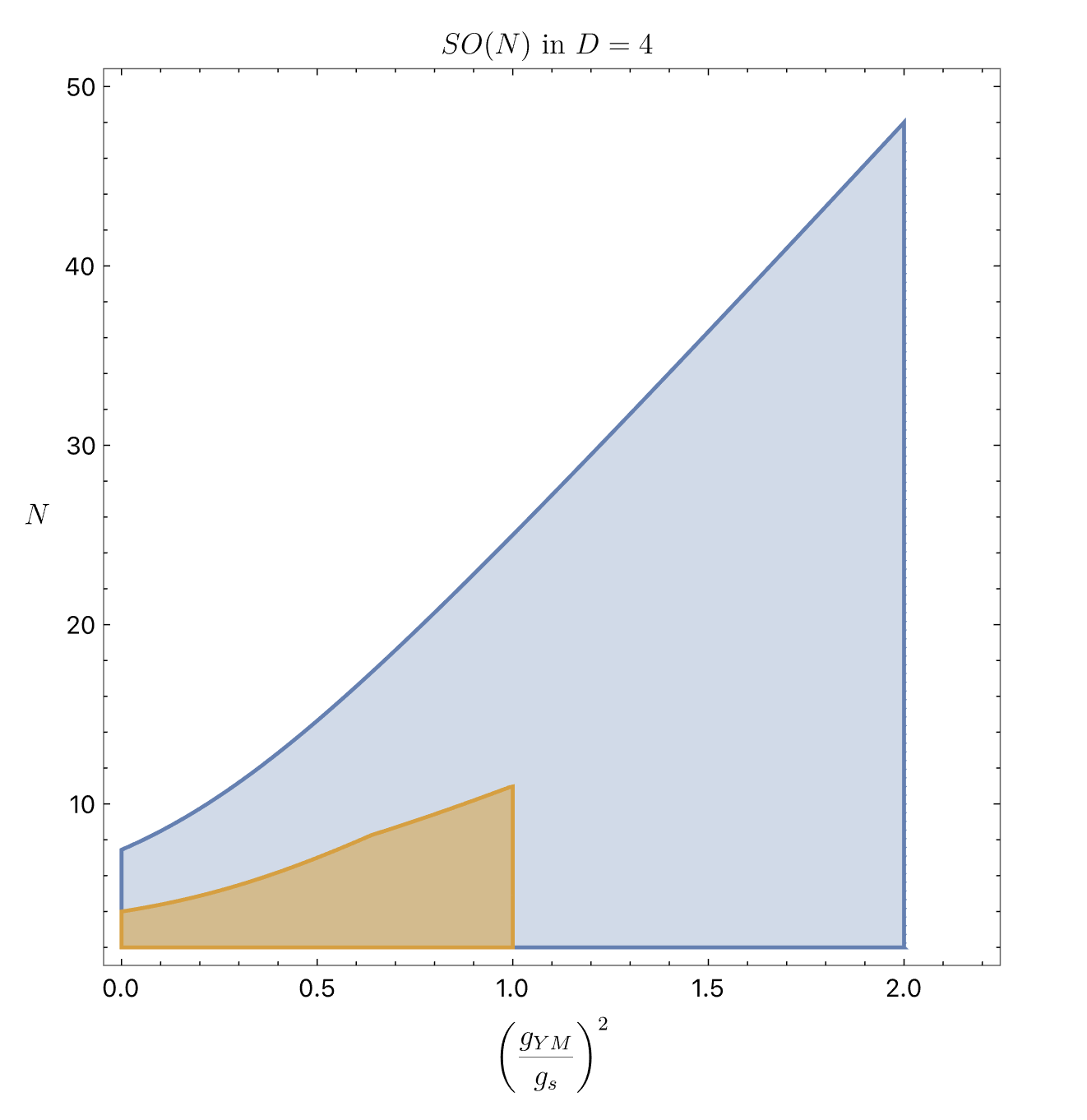}\hspace{11mm} \includegraphics[scale=.4]{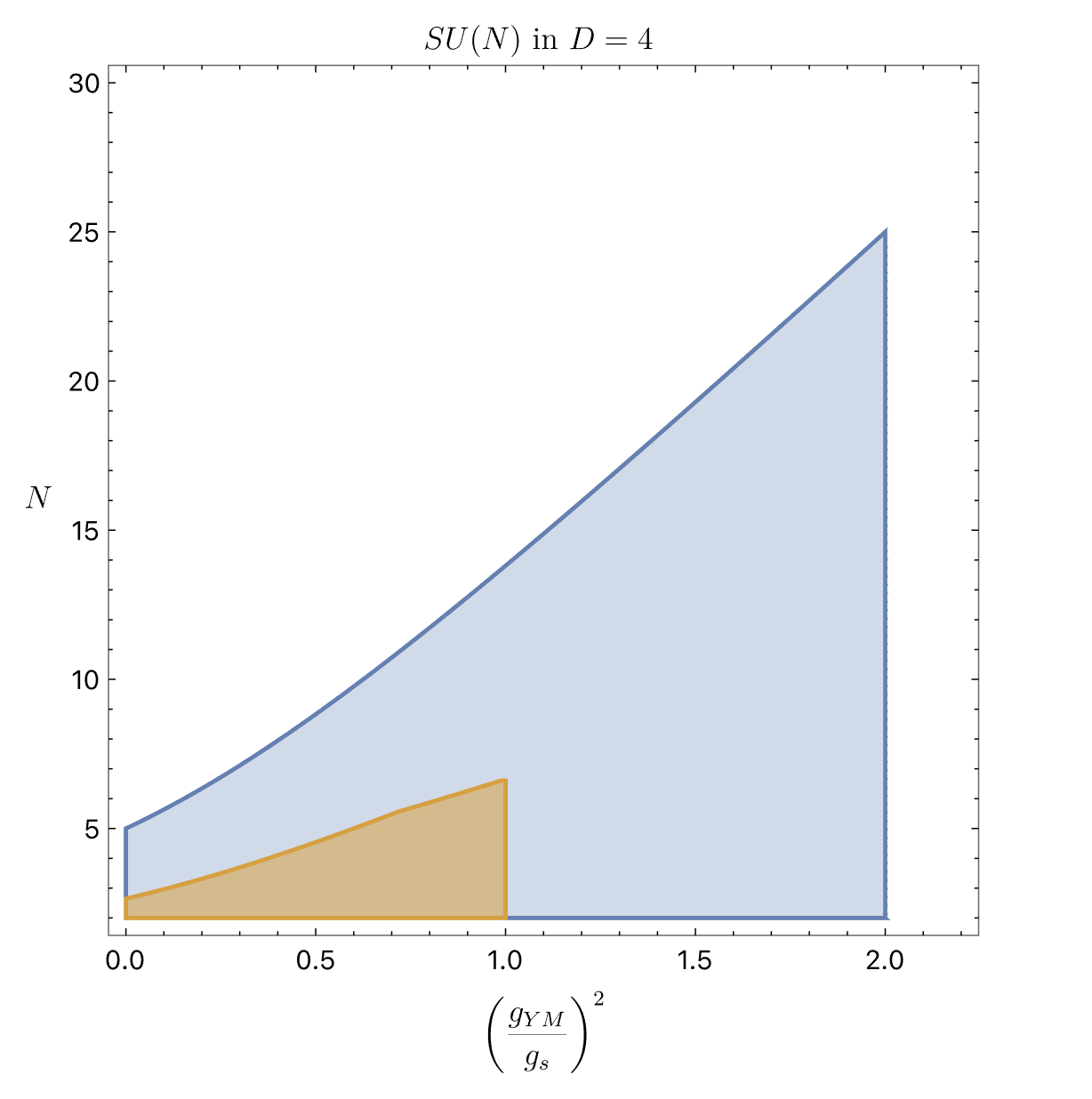}
	\caption{Allowed space of $N$ and $g_{YM}/g_s$ in four spacetime dimensions, for $SO(N)$ (left) and $SU(N)$ (right).  In both cases, the larger allowed region (blue) has heterotic poles and contains the region without heterotic poles (orange).  In the heterotic case, the maximum rank is $24$ for both groups and occurs at the coupling $g_{YM}^2 = 2g_s^2$, the value fixed in the heterotic string.}
	\label{fig:sonsun4d}
\end{figure}
\paragraph{SO(N):} For $SO(N)$ the adjoints exchange six representations.  Positivity of $g_{YM}^2$ is already a consistency condition imposed at massless level.  As can be seen in the left plot of figure \ref{fig:sonsun4d}, both the heterotic and non-heterotic cases have an $N$-independent upper bound on the ratio $\frac{g_{YM}^2}{g_s^2}$ and an upper bounding curve depending on both the coupling and $N$.  In the non-heterotic case this upper curve is only piecewise smooth, the result of two smooth constraints.  In particular, we find for the non-heterotic case, the constraints and the associated mass-level, spin, and $SO(N)$ representation are the following:
\begin{equation}
	\begin{cases}
		\begin{tikzpicture}
	\node at (0, 0) {\includegraphics[scale=.6]{figpdfs/sonbig-figure0.pdf} };
	\node at (1.2, -1 ) {\tiny $m^2 = 1, J = 0$};
	\node at (5, 0) {$g_{YM}^2 \leq g_s^2$};
\end{tikzpicture}\\\\
\hspace{11mm}\begin{tikzpicture}
	\draw[fill] (2, -.5) circle (1mm);
	\node at (3, -1 ) {\tiny $m^2 = 1, J = 0$};
	\node at (7.75, -.5) {$1+\left(\frac{g_{YM}}{g_s} \right)^2( N-2)-\frac{N(N-1)}{12} \geq 0$};
\end{tikzpicture}\\\\
\hspace{11mm}\begin{tikzpicture}
	\draw[fill] (2, -.5) circle (1mm);
	\node at (3, -1 ) {\tiny $m^2 = 2, J = 0$};
	\node at (7.75, -.5) {$4+2\left(\frac{g_{YM}}{g_s} \right)^2( N-2)-\frac{N(N-1)}{5} \geq 0$};
\end{tikzpicture}
	\end{cases}
\end{equation}
And in the heterotic case simply the two constraints:
\begin{equation}
	\begin{cases}
		\begin{tikzpicture}
	\node at (0, 0) {\includegraphics[scale=.6]{figpdfs/sonbig-figure0.pdf} };
	\node at (1.2, -1 ) {\tiny $m^2 = 1, J = 0$};
	\node at (5, 0) {$g_{YM}^2 \leq 2g_s^2$};
\end{tikzpicture}\\\\
\hspace{11mm}\begin{tikzpicture}
	\draw[fill] (2, -.5) circle (1mm);
	\node at (3, -1 ) {\tiny $m^2 = 1, J = 0$};
	\node at (7.75, -.5) {$2+\left(\frac{g_{YM}}{g_s} \right)^2( N-2)-\frac{N(N-1)}{24} \geq 0$};
\end{tikzpicture}\\
	\end{cases}
\end{equation}
Where the masses $m^2$ are measured in units of $M_s^2$.  The two-by-two tableaux is the largest representation exchanged between the gauge bosons, having dimension $\frac{N(N+1)(N+2)(N-3)}{12}$, and is the exchanged state enforcing the anti-weak-gravity bound.  We note that all constraints come from the most subleading Regge trajectory.
\paragraph{SU(N):} For $SU(N)$ the adjoints exchange seven representations, bu the situation is analogous, with analogous representations imposing similar bounds.  We find for the non-heterotic case: 
\begin{equation}
	\begin{cases}
		\begin{tikzpicture}
	\node at (0, 0) {\includegraphics[scale=.7]{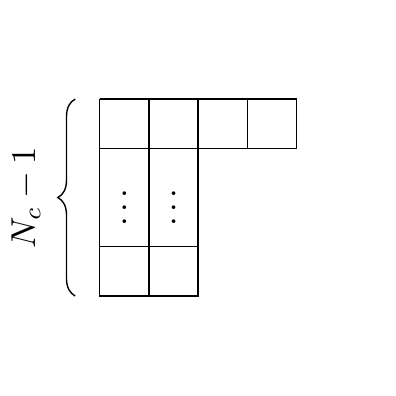} };
	\node at (1.2, -1 ) {\tiny $m^2 = 1, J = 0$};
	\node at (5, 0) {$g_{YM}^2 \leq g_s^2$};
\end{tikzpicture}\\\\
\hspace{11mm}\begin{tikzpicture}
	\draw[fill] (2, -.5) circle (1mm);
	\node at (3, -1 ) {\tiny $m^2 = 1, J = 0$};
	\node at (7.75, -.5) {$5+4\left(\frac{g_{YM}}{g_s} \right)^2N-\frac{1+2N^2}{3} \geq 0$};
\end{tikzpicture}\\\\
\hspace{11mm}\begin{tikzpicture}
	\draw[fill] (2, -.5) circle (1mm);
	\node at (3, -1 ) {\tiny $m^2 = 2, J = 0$};
	\node at (7.75, -.5) {$11+5\left(\frac{g_{YM}}{g_s} \right)^2N-N^2\geq 0$};
\end{tikzpicture}
	\end{cases}
\end{equation}
and in the heterotic case:
\begin{equation}
	\begin{cases}
		\begin{tikzpicture}
	\node at (0, 0) {\includegraphics[scale=.7]{figpdfs/sunbig-figure0.pdf} };
	\node at (1.2, -1 ) {\tiny $m^2 = 1, J = 0$};
	\node at (5, 0) {$g_{YM}^2 \leq 2g_s^2$};
\end{tikzpicture}\\\\
\hspace{11mm}\begin{tikzpicture}
	\draw[fill] (2, -.5) circle (1mm);
	\node at (3, -1 ) {\tiny $m^2 = 1, J = 0$};
	\node at (7.75, -.5) {$\left(\frac{g_{YM}}{g_s} \right)^2N-\frac{1}{12} (N^2-25) \geq 0$};
\end{tikzpicture}
	\end{cases}
\end{equation}
We see that for $SU(N)$ again the largest representation enforces an identical bound on the coupling to the $SO(N)$ case, this time he dimension of the representation is $\frac{N^2(N+3)(N-1)}{4}$.  These  bounds are not the same function of rank $r$ for each group, but the bound on the rank is identical at the maximum value of $g_{YM}^2$ in the heterotic case, which is $2g_s^2 = \frac{2M_s^2}{M_P^2} $.  This is the condition relating the Yang-Mills coupling, the string scale, and the Planck scale in the heterotic string, a condition which survives compactification as both ten-dimensional couplings get the same volume dilution \cite{Gross:1985rr}. 
\paragraph{Phenomenology} Here we comment on the implications for four-dimensional phenomenology.  First we note the obvious, which is that only the combination of constraints from singlets of the gauge group, massless adjoints, and the largest exchanged representation both for $SO(N)$ and $SU(N)$ combine to produce a bounded region.  The largest representation furnishes the bound relating the UV completion scale, Planck scale, and the coupling:
\begin{equation}
	\label{eq:antiwgc}
	g_{YM}^2 M_P^2 \leq  2M_s^2
\end{equation}
or with no factor of two on the right-hand side when we do not have heterotic denominators.  This is an anti-weak-gravity type of bound: consistent with weak-gravity, but pointing in the opposite direction. But for sufficiently large ranks of the gauge-group, we cannot afford for the gauge-coupling to be too-weak with a low string scale, either.  In particular, no global symmetry is allowed for these massless adjoint vectors for $N > 7$ for $SO(N)$ and $N > 5$ for $SU(N)$.\\
We can discuss these constraints in the context of $W$ bosons and gluons.  For the case of heterotic poles, the constraint is the same for both $W$'s and gluons and is (\ref{eq:antiwgc}).  For couplings around the GUT scale, we have $g_2^2$, $g_3^2 \sim \frac{1}{2}$, in which case the putative string scale is bound by 
\begin{equation}
	\frac{M_P}{2} \lesssim M_s
\end{equation}
meaning the lowest putative string scale is around $10^{18}$ GeV.  For the case without heterotic poles we see that the constraint on $M_s^2$ changes by a factor of two.  Though there is also a lower bound for $SU(3)$ in the non-heterotic case, the bound is not interesting unless the Yang-Mills coupling is sufficiently weak.  Otherwise, the constraint $g_s < 1 $ justifying the perturbative analysis is much stronger anyway.  Nonetheless, for sufficiently weak $g_{YM}^2$ (of order a tenth) we see that we get a two-sided bound on $M_s$.  It is also amusing to note that in the case of non-heterotic poles, $SU(5)$ and $SO(10)$ gauge bosons are essentially pegged to have $g_{YM}^2 = 2g_s^2$.
\subsection{Constraints in General Spacetime Dimensions}
In general dimensions, the full constraints require $D$-dimensional spinning polynomials.  The Yang-Mills amplitude in $D$ spacetime dimensions is of the form 
\begin{equation}
	\mathcal{A} = \mathcal{F}^4 \mathcal{A}^{\text{scalar}}
\end{equation}
where $\mathcal{A}^{\text{scalar}}$ has no dependence on polarization vectors and $\mathcal{F}^4$ is the famous polynomial permutation-invariant in field strengths which sits in front of the field theory Yang-Mills amplitude.  In \cite{unitaritystringamps}, positivity of $\mathcal{F}^4$ alone was remarkably shown to contain the critical dimension constraint $D \leq 10$ and information about the spectrum of 11-dimensional supergravity.  This is the only condition for positive expandability of $\mathcal{F}^4$.  For our analysis, what is relevant is that so long as we satisfy the critical dimension constraint $D \leq 10$, positive expandability of $\mathcal{A}^{\text{scalar}}$ on the scalar $D$-dimensional Gegenbauer polynomials then implies positive expandability of the full amplitude's residues, as we merely have a product of two positively expandable functions, which in turn must be positively expandable.  So for general dimensions equal to or below ten, we will consider the sufficient though not strictly necessary condition of positive expandability on the \textit{scalar} $D$-dimensional Gegenbauer polynomials.  This means that the true constraints could in principle be weaker, but merely expanding on the scalar polynomial basis already provides interesting constraints. So we have\footnote{These would be denoted in the math literature as $C^{(\frac{D-3}{2} )}_J(\cos\theta)$}
\begin{equation}
	\label{eq:spinninggegs}
	\mathbb{G}_{D, J}^{\{h_i\}}(\cos\theta) = G^{(D)}_{J}(\cos\theta)
\end{equation}
It is worth commenting that in four spacetime dimensions, where we did the full spinning analysis, the strongest constraints for residues in the $s$-channel come from the process with $h_1 = h_2 = -1$ and $h_3 = h_4 = +1$ for which the spinning polynomials reduce to the Legendre polynomials. This might hint at the constraints on $\mathcal{A}^{\text{scalar}}$ constituting the full set of constraints, but we will leave such analysis to future work.
\paragraph{Heterotic $SO(N)$ in Ten Dimensions}
\begin{figure}[h!]
\centering
  \includegraphics[scale=.5]{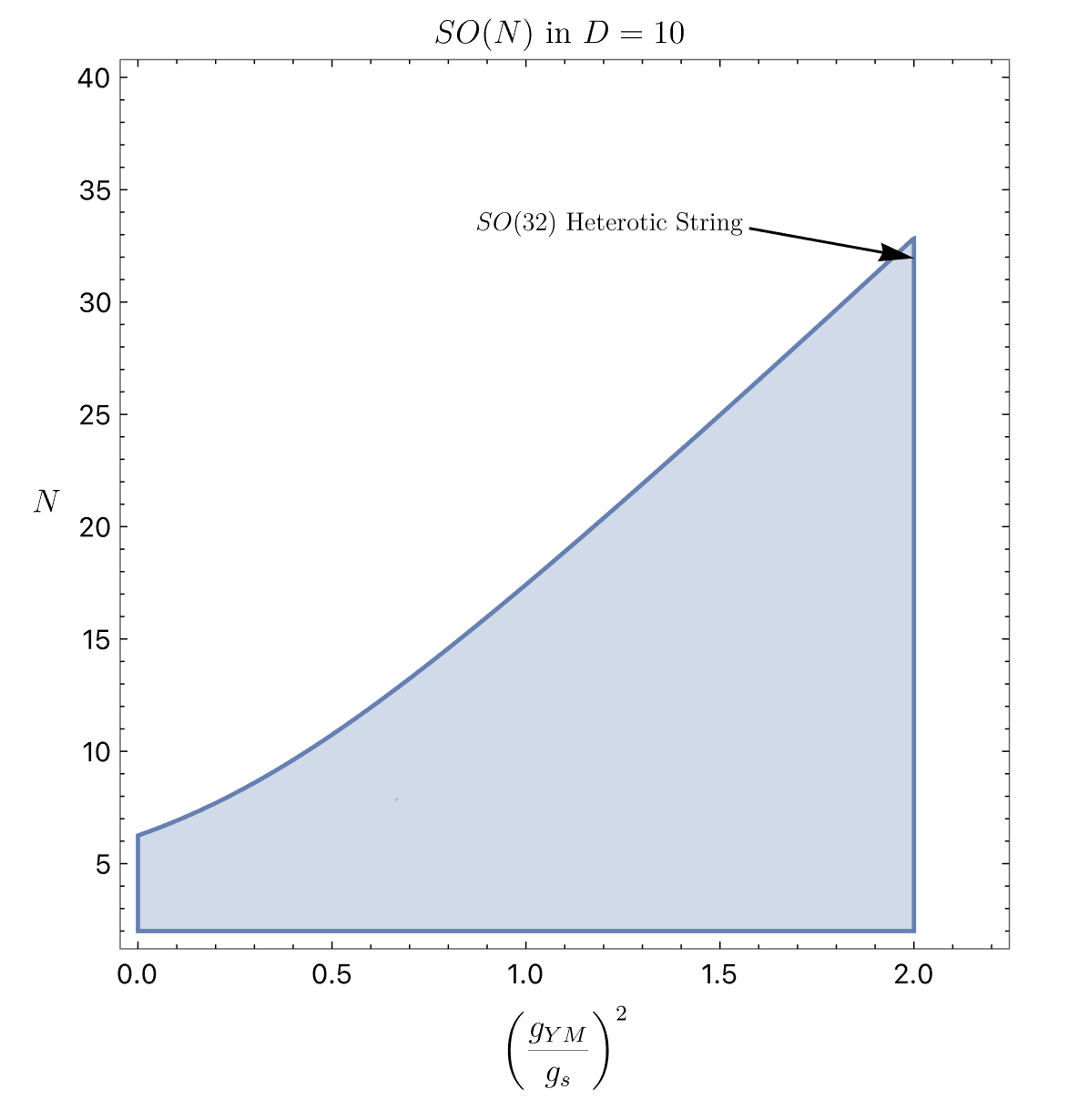} 
  \caption{Imposing positive expandability of the heterotic form of (\ref{eq:fourgaugeboson}) on scalar Gegenbauer's in $D = 10$, we find the allowed region shaded in blue.  The value of coupling and $N$ fixed in $SO(32)$ heterotic string theory is indicated by the arrow, $g_{YM}^2 = 2g_s^2$ and $N = 32$.}
  \label{fig:so32}
\end{figure}
Given that our ansatz is essentially the heterotic string amplitude (the gauge group and gauge-coupling are not fixed), it is natural to check the constraints of the previous section but in ten dimensions, which can be found in figure \ref{fig:so32}.  Similar exchanged states produce the analogous curves in the $D = 10$ case, the only difference being that the strongest upper bounding curve occurs at mass level three rather than one.  In particular, the maximum allowed value of the coupling ratio $\frac{g_{YM}^2}{g_s^2} $ is 2, enforced by the same exchanged state as in four dimensions, and this meets the upper bounding curve from mass-level three singlet exchange at the corner of the allowed region with the maximum allowed
\begin{equation}
	\label{eq:sonbound}
	N \leq 32+\frac{16}{19}
\end{equation}
which of course means the only theory allowed by this corner of the allowed region is the $SO(32)$ heterotic string.  
\paragraph{Dimension vs. Rank}
In general dimensions we observe that the maximum allowed rank of the gauge group always occurs at $g_{YM}^2 = 2 g_s^2$, the maximum allowed value of the coupling and also the value in the $SO(32)$ heterotic string.  We can then track how the maximum allowed rank varies with the spacetime dimension, which is presented in figure \ref{fig:rankvsd}.
\begin{figure}[h!]
\flushleft
	\begin{tikzpicture}[scale=1]
		\node at (0, 0) {\includegraphics[scale=.485]{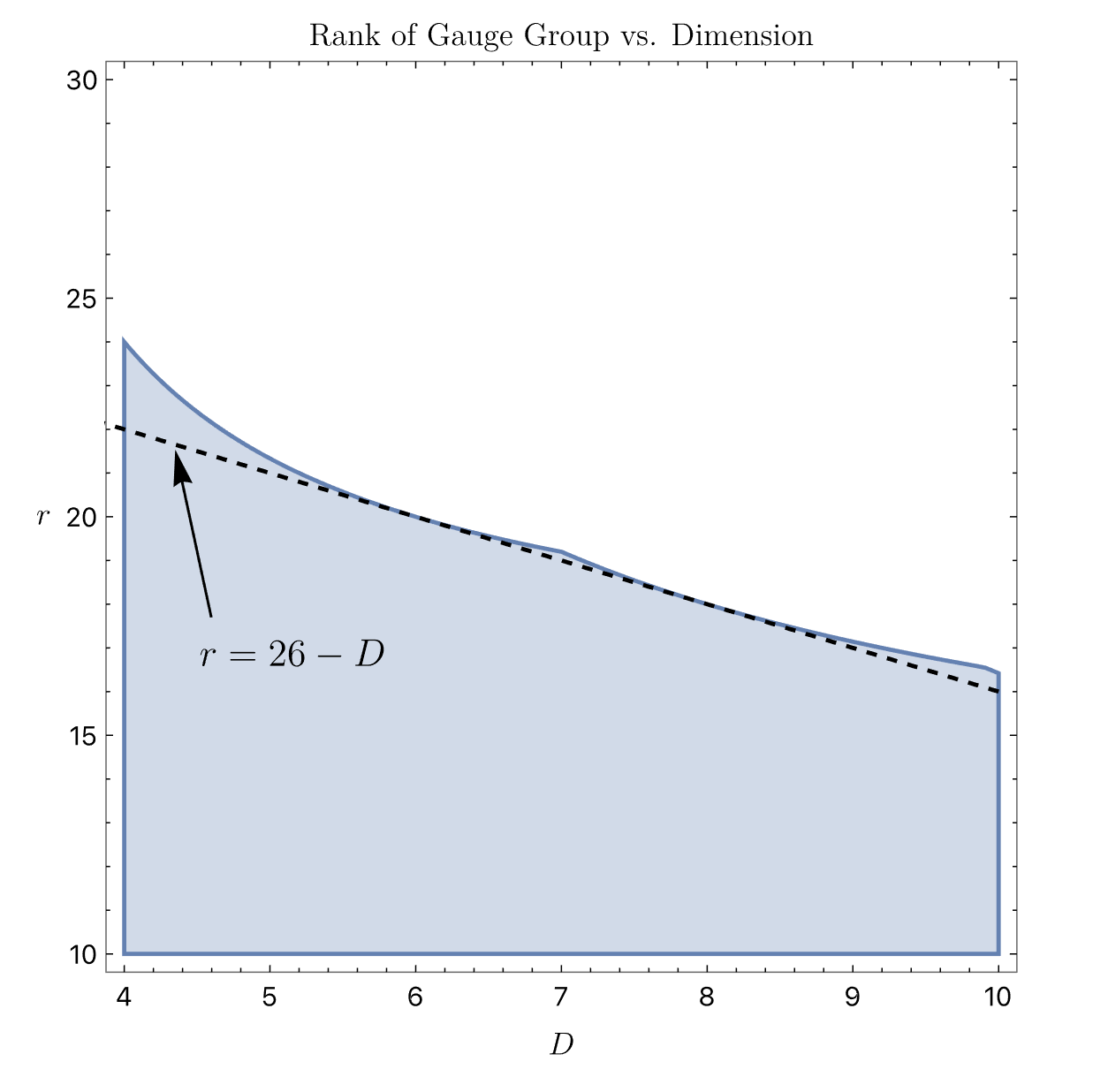}};
		\node at (9.5, 2.25) {\includegraphics[scale=.25]{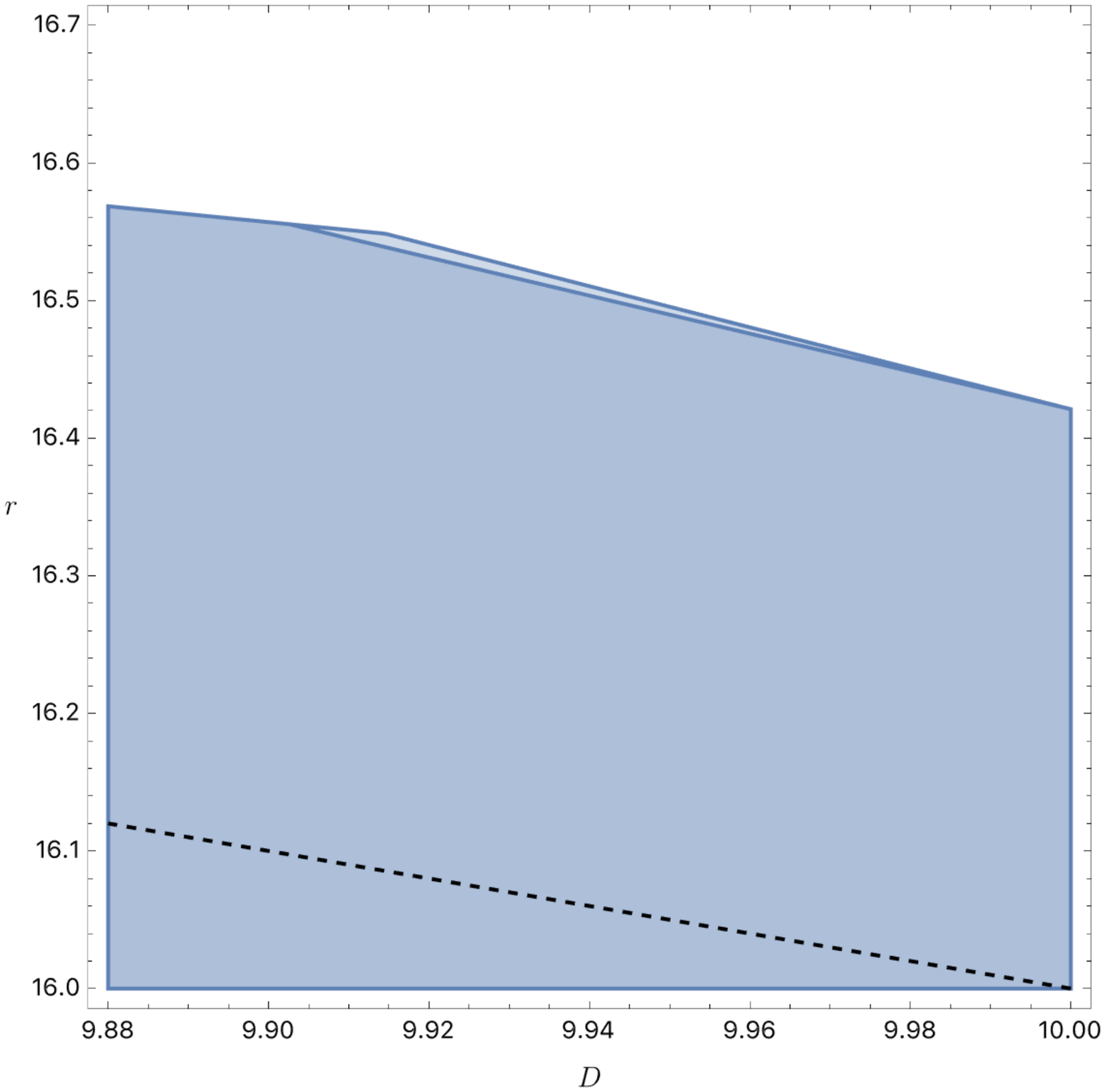}};
		\draw (4.5, -1.45) rectangle (4, -.95);
		\draw (4.5, -.95) -- (7, -.125);
		\node at (10.75, 4.35) {\tiny$SO(N)$};
		\draw[->] (10.25, 4.25) -- (8.75, 3.75);
		%\node at (0.5, 5.5) {\text{Rank vs. Dimension}}; 
	\end{tikzpicture}
	\vspace{-10mm}\caption{Imposing positive expandability of the heterotic form of (\ref{eq:fourgaugeboson}) on scalar $D$-dimensional Gegenbauer's with $g_{YM}^2 = 2g_s^2$, we find the allowed region in rank $r$ vs. dimension $D$ shaded in blue.  There are three-piecewise segments, the two leftmost are identical for $SO(N)$ and $SU(N)$.  The zoomed portion shows that in the vicinity of $D = 10$, the third curve is not the same curve in $r$ and $D$ for $SO(N)$ and $SU(N)$ but agrees in $D = 10$, the constraints agreeing in all physical dimensions. }
	\label{fig:rankvsd}
\end{figure}
We look at bounds between four and ten dimensions, over which three distinct smooth bounding curves comprise the full upper bounding curve.  The bounding curve always corresponds to a spin-0 singlet of the gauge group, the only quantum number varying being the mass.  For dimensions $4 \leq D \leq 7$ the constraint comes from the first mass level.  In higher dimensions the strongest constraint comes from the second mass level until in the vicinity of $D = 10$, at which a yet stronger constraint comes in from the third mass level.  This is seen in the zoomed portion of figure \ref{fig:rankvsd}.  In $D = 10$ this constraint from mass level three is necessary to disallow $N = 33$ in the case of $SO(N)$. The dashed line on figure \ref{fig:rankvsd} represents the swampland conjecture $r < 26-D$, derived assuming BPS completeness with 16 supercharges \cite{Kim:2019ths} .  We emphasize that the first two of these upper bounding curves is identical for $SO(N)$ and $SU(N)$, the agreement being not at all manifest even at the level of the analytic expressions in $r$ and $D$ which are required to be positive; the expressions merely share a common factor enforcing positivity.  Even more dramatically, the zoomed portion reveals the constraint at the third mass level is not even the same curve for $SU(N)$ and $SO(N)$, but the two curves are such that they meet at $D = 10$ constraining the rank to be $r \leq 16+\frac{8}{19}$, so the rank constraint in any physical dimension is identical.  This non-trivial conspiracy of the Gegenbauer polynomials and recoupling coefficients suggests a universality to these constraints when centered on the correct data which could perhaps be exploited in greater generality.
\paragraph{Swampland Conjectures:} In the context of these ans{\"a}tze we are able to make contact with some swampland conjectures, in particular weak gravity and consequences of completeness, such as the conjecture about the maximum allowed rank of the gauge group.  In the context of gravitational completions, we also see the mechanism for generating something like completeness.  Placing the field theory amplitude in front of a common stringy form-factor means that the $u$ and $t$ channel projectors for the gauge-group must be re-expanded in terms of the $s$-channel projectors when we take a residue in $s$.  This is done via the group theory crossing equation which is solved via 6-$j$ symbols.  In this way, we see that scattering some specified representations builds up the need for other representations in their tensor product.  This kind of mechanism for completeness has everything to do with gravity, as one can note that with the non-gravitational NLSM completion (\ref{eq:ampnlsmUV}) the poles merely produce the singlet and adjoint exchanges.  Even for a stringy completion of the NLSM amplitude via Lovelace-Shapiro, we still only generate either anti-symmetric adjoint exchange or symmetric adjoint and singlet exchange.  But in the context of our gravitational amplitudes, we find every possible representation that can be exchanged between the external states is indeed exchanged.  This provides a mechanism to bootstrap the completeness hypothesis in the context of these amplitudes, one which crucially relies on the presence of gravity.
\section{Standard Model Electroweak Sector}
\label{sec:SM}
Now we direct our attention to the electroweak sector.  We already studied ans{\"a}tze for the scattering of $SU(N)$ gauge bosons, and found a constraint on the relation between the gauge-coupling, the UV completion scale, and the Planck scale. The only constraint coming from scattering $SU(N)$ gauge bosons in the heterotic case with $N \leq 5$ was 
\begin{equation}
	g_{YM}^2 M_P^2 \leq 2M_s^2
	\end{equation}
 Then the lowest string scale we can obtain is 
 \begin{equation}
 	M_s^2 \sim  10^{18} \text{ GeV}
 \end{equation}
and the bound pushes $M_s^2$ up by a factor of two in the non-heterotic case.
 We can further probe the electroweak sector by studying the scattering of four Higgs's.  The amplitude in this case is 
 \begin{multline}
 \label{eq:fourhiggsamp}
 	\mathcal{A}(1, \bar 2, 3, \bar 4) = \Gamma^{\text{str}}\bigg(-\frac{1}{M_P^2}\left( \frac{tu}{s}\mathbb{P}_{1, 1}^s+\frac{su}{t}\mathbb{P}_{1,1}^t\right)+\frac{t-u}{2s}\left(\frac{g_1^2}{4} \mathbb{P}_{1, 1}^s+g_2^2\mathbb{P}_{\text{Adj}, 1}^s\right)\\+\frac{s-u}{2t} \left(\frac{g_1^2}{4} \mathbb{P}_{1, 1}^t+g_2^2\mathbb{P}_{\text{Adj}, 1}^t\right)
 	+2\lambda (\mathbb{P}_{1,1}^s+\mathbb{P}_{1,1}^t) \bigg)
 \end{multline}
and similarly for the configuration with massless $t$ and $u$ channel poles only.  The subscript labels on the projector denote the exchanged representation of the corresponding factor of $SU(2)\times U(1)_Y$ with $1$ denoting singlet exchange (with zero charge in the $U(1)_Y$ case).  The projectors are normalized such that these are the conventional normalizations for the gauge-couplings and Higgs quartic coupling in the standard model.  If we fix the gauge-couplings such that $\alpha_S^{-1} = \alpha_W^{-1} = 25$ we can produce a plot relating the Higgs quartic coupling and the putative string scale in Planck units.
\begin{figure}[h!]
\centering
	\includegraphics[scale=.635]{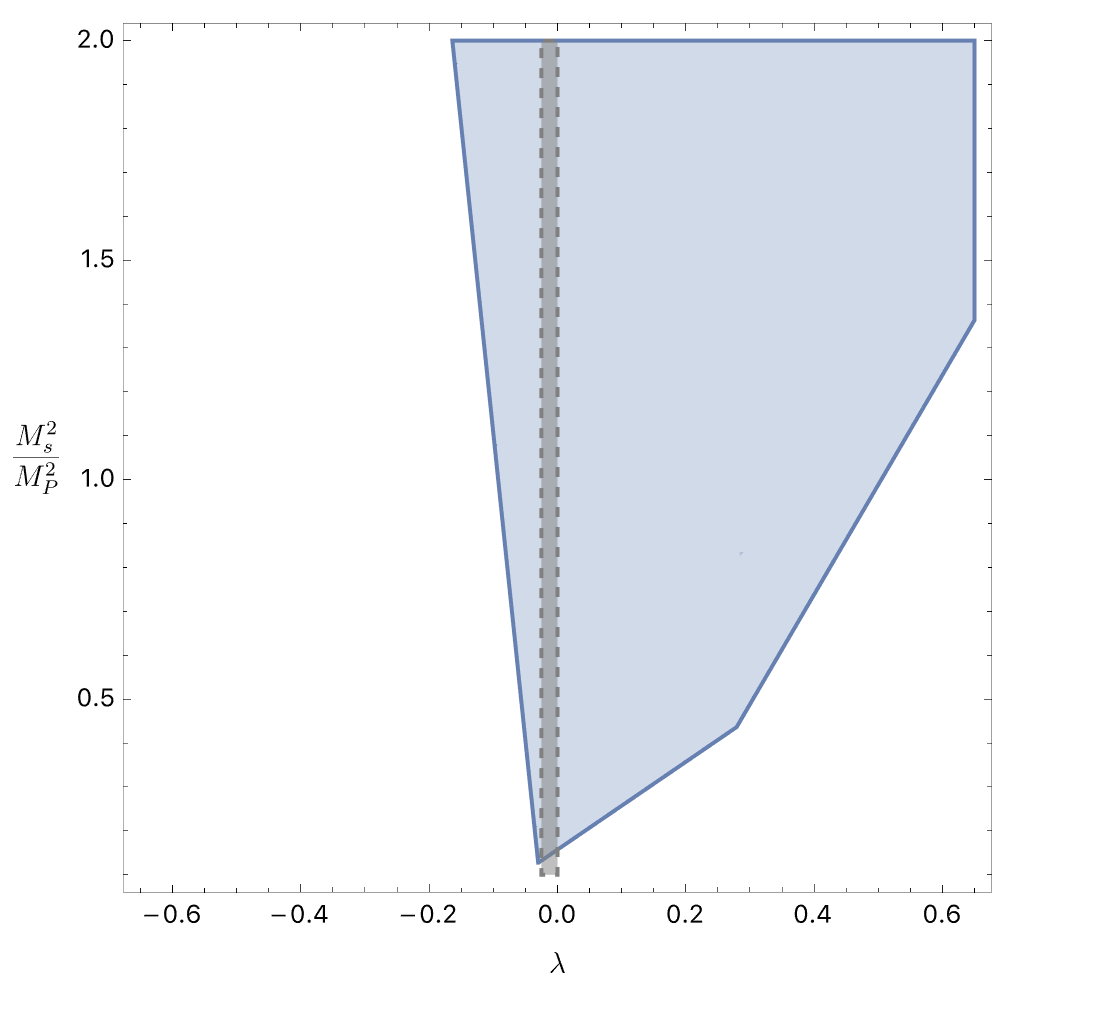}
	\caption{Plot of allowed region (blue) for Higgs quartic coupling $\lambda$ versus the string scale squared $M_s^2$ in reduced Planck units.  The dashed lines bounding the shaded green band are roughly $3\sigma$ bounds for the SM running of $\lambda$ with uncertainty coming from the uncertainty in the mass of the top quark (taken from \cite{Degrassi:2012ry}).  The bounds come from imposing unitarity of (\ref{eq:fourhiggsamp}).  With heterotic gravitational denominators, $\lambda$ is strictly positive.}
	\label{fig:higgsquartic}
\end{figure}
The kink in the allowed region minimizing $M_s$ is just outside of the bounds of the running of $\lambda$ predicted by the standard model \cite{Degrassi:2012ry}.  The running predicted has $\lambda$ become negative at $\sim 10^{10}\text{ GeV}$ and asymptote to within the gray shaded region of figure \ref{fig:higgsquartic} below the Planck scale.  The lowest scale allowed by this bound is slightly below that allowed by the four gauge boson scattering, constituting a weaker constraint.\\
It is crucial to note though, that we find this bound on the Higgs quartic coupling in the absence of heterotic gravitational denominators.  If instead we have the heterotic denominators, then we find that $\lambda$ is strictly positive, with the explicit lower bound varying with the gauge couplings.  It is also worth noting that in analogy with these poles being associated with gauge bosons' and graviton's non-minimal couplings to the dilaton via $F^2\phi$ and $R^2 \phi$ in the heterotic string, we would expect a coupling $(H^\dag H)\phi$ for these amplitudes.   In the context of many BSM models, the Higgs quartic running is modified by new states below $\sim 10^{10} \text{GeV}$, and could be consistent with the small positive $\lambda$'s allowed by this heterotic analogue of (\ref{eq:fourhiggsamp}).  But this is at least naively in tension with the only new states coming in at this putative string scale, a scale which gauge-boson scattering required be only slightly below the Planck scale for $\alpha$'s $\sim \frac{1}{25}$.  This basic tension in the $2 \to 2$ Higgs amplitude already poses an interesting puzzle in this particular bottom up approach to building stringy completions of the standard model.    
\section{Conclusions Outlook}
In this work we considered a class of stringy ans{\"a}tze  for $2 \to 2$ amplitudes and constrained them via unitarity.  The stringy nature of these amplitudes flipped gravity's role from an obstruction to UV softening to a necessary condition for UV softening consistent with positive expandability of the imaginary part, with constraints bounding the strength of other interactions relative to gravity from above.  We found that gauge boson scattering in particular furnished stringent constraints, especially in higher dimensions, where we found that the $SO(32)$ heterotic string is barely consistent with  perturbative unitarity.  Perhaps most compelling is the agreement between the suggested constraints on rank versus spacetime dimension for both $SO(N)$ and $SU(N)$ gauge boson scattering not only with each other but with the swampland conjecture $r < 26-D$ in greater than four dimensions.  These results indicate that perturbative unitarity constraints can become particularly potent in the context of gravity.  There are a variety of further questions to pursue.\\
These amplitudes morally resolved the particles into closed strings. It would be interesting to pursue the open string analogue of this analysis which would correspond to braneworld scenarios.  In this case the leading unitarizing interactions are the gauge-boson exchanges, not gravity, and a genus 0 analysis will not furnish constraints on gauge or global symmetry groups, though one can still expect constraints on couplings.  In order to get gravity in the low-energy limit, which makes symmetry constraints more likely, we will need control over ans{\"a}tze at genus 1.\\
Another interesting extension of this work would be considering deformations like the Coon amplitude or the more general deformations considered in \cite{Cheung:2022mkw}.  Again here, deformations of closed string amplitudes will provide more readily amenable to this analysis as they will provide generalizations of the above constraints already at genus 0.\\
Finally, in order to continue to probe the consistency of the amplitudes considered herein, two obvious avenues are to develop the analysis both for higher points and for massive external legs.  The analysis above not only obtains bounds from requiring the exchange of positive norm states, but tells you the quantum numbers of such states.  One can iteratively build up amplitudes with these massive states as external legs and impose further consistency.
\paragraph{Acknowledgments} We would like thank Wayne Zhao for initial collaboration.  We would also like to thank Sebastian Mizera, Lorenz Eberhardt, and Yu-tin Huang for useful discussions. And we would like to especially thank Nima Arkani-Hamed for many valuable discussions and comments on the draft. 

\bibliographystyle{utphys}
{\linespread{1.075}
\bibliography{draftv4.bib}

\providecommand{\href}[2]{#2}\begingroup\raggedright\begin{thebibliography}{10}

\bibitem{Virasoro:1969me}
M.~A. Virasoro, ``{Alternative constructions of crossing-symmetric amplitudes
  with regge behavior},''
  \href{http://dx.doi.org/10.1103/PhysRev.177.2309}{{\em Phys. Rev.} {\bfseries
  177} (1969) 2309--2311}.

\bibitem{Cvitanovic:2008zz}
P.~Cvitanovic, {\em {Group theory: Birdtracks, Lie's and exceptional groups}}.
\newblock 2008.

\bibitem{Gross:1985rr}
D.~J. Gross, J.~A. Harvey, E.~J. Martinec, and R.~Rohm, ``{Heterotic String
  Theory. 2. The Interacting Heterotic String},''
  \href{http://dx.doi.org/10.1016/0550-3213(86)90146-X}{{\em Nucl. Phys. B}
  {\bfseries 267} (1986) 75--124}.

\bibitem{unitaritystringamps}
N.~Arkani-Hamed, L.~Eberhardt, Y.-t. Huang, and S.~Mizera, ``{On unitarity of
  tree-level string amplitudes},''
  \href{http://dx.doi.org/10.1007/JHEP02(2022)197}{{\em JHEP} {\bfseries 02}
  (2022) 197}, \href{http://arxiv.org/abs/2201.11575}{{\ttfamily
  arXiv:2201.11575 [hep-th]}}.

\bibitem{Kim:2019ths}
H.-C. Kim, H.-C. Tarazi, and C.~Vafa, ``{Four-dimensional
  $\mathbf{\mathcal{N}=4}$ SYM theory and the swampland},''
  \href{http://dx.doi.org/10.1103/PhysRevD.102.026003}{{\em Phys. Rev. D}
  {\bfseries 102} no.~2, (2020) 026003},
  \href{http://arxiv.org/abs/1912.06144}{{\ttfamily arXiv:1912.06144
  [hep-th]}}.

\bibitem{Degrassi:2012ry}
G.~Degrassi, S.~Di~Vita, J.~Elias-Miro, J.~R. Espinosa, G.~F. Giudice,
  G.~Isidori, and A.~Strumia, ``{Higgs mass and vacuum stability in the
  Standard Model at NNLO},''
  \href{http://dx.doi.org/10.1007/JHEP08(2012)098}{{\em JHEP} {\bfseries 08}
  (2012) 098}, \href{http://arxiv.org/abs/1205.6497}{{\ttfamily arXiv:1205.6497
  [hep-ph]}}.

\bibitem{Cheung:2022mkw}
C.~Cheung and G.~N. Remmen, ``{Veneziano Variations: How Unique are String
  Amplitudes?},'' \href{http://arxiv.org/abs/2210.12163}{{\ttfamily
  arXiv:2210.12163 [hep-th]}}.

\bibitem{NimaEFT}
N.~Arkani-Hamed, T.-C. Huang, and Y.-T. Huang, ``{The EFT-Hedron},''
  \href{http://dx.doi.org/10.1007/JHEP05(2021)259}{{\em JHEP} {\bfseries 05}
  (2021) 259}, \href{http://arxiv.org/abs/2012.15849}{{\ttfamily
  arXiv:2012.15849 [hep-th]}}.

\bibitem{Caron-Huot:2022jli}
S.~Caron-Huot, Y.-Z. Li, J.~Parra-Martinez, and D.~Simmons-Duffin, ``{Graviton
  partial waves and causality in higher dimensions},''
  \href{http://arxiv.org/abs/2205.01495}{{\ttfamily arXiv:2205.01495
  [hep-th]}}.

\bibitem{Caron-Huot:2022ugt}
S.~Caron-Huot, Y.-Z. Li, J.~Parra-Martinez, and D.~Simmons-Duffin, ``{Causality
  constraints on corrections to Einstein gravity},''
  \href{http://arxiv.org/abs/2201.06602}{{\ttfamily arXiv:2201.06602
  [hep-th]}}.

\bibitem{Bern:2021ppb}
Z.~Bern, D.~Kosmopoulos, and A.~Zhiboedov, ``{Gravitational effective field
  theory islands, low-spin dominance, and the four-graviton amplitude},''
  \href{http://dx.doi.org/10.1088/1751-8121/ac0e51}{{\em J. Phys. A} {\bfseries
  54} no.~34, (2021) 344002}, \href{http://arxiv.org/abs/2103.12728}{{\ttfamily
  arXiv:2103.12728 [hep-th]}}.

\bibitem{Guerrieri:2020bto}
A.~L. Guerrieri, J.~Penedones, and P.~Vieira, ``{S-matrix bootstrap for
  effective field theories: massless pions},''
  \href{http://dx.doi.org/10.1007/JHEP06(2021)088}{{\em JHEP} {\bfseries 06}
  (2021) 088}, \href{http://arxiv.org/abs/2011.02802}{{\ttfamily
  arXiv:2011.02802 [hep-th]}}.

\bibitem{Guerrieri:2021ivu}
A.~Guerrieri, J.~Penedones, and P.~Vieira, ``{Where Is String Theory in the
  Space of Scattering Amplitudes?},''
  \href{http://dx.doi.org/10.1103/PhysRevLett.127.081601}{{\em Phys. Rev.
  Lett.} {\bfseries 127} no.~8, (2021) 081601},
  \href{http://arxiv.org/abs/2102.02847}{{\ttfamily arXiv:2102.02847
  [hep-th]}}.

\bibitem{Albert:2022oes}
J.~Albert and L.~Rastelli, ``{Bootstrapping pions at large N},''
  \href{http://dx.doi.org/10.1007/JHEP08(2022)151}{{\em JHEP} {\bfseries 08}
  (2022) 151}, \href{http://arxiv.org/abs/2203.11950}{{\ttfamily
  arXiv:2203.11950 [hep-th]}}.

\bibitem{Soldate:1986mk}
M.~Soldate, ``{Partial Wave Unitarity and Closed String Amplitudes},''
  \href{http://dx.doi.org/10.1016/0370-2693(87)90302-9}{{\em Phys. Lett. B}
  {\bfseries 186} (1987) 321--327}.

\bibitem{Arkani-Hamed:2006emk}
N.~Arkani-Hamed, L.~Motl, A.~Nicolis, and C.~Vafa, ``{The String landscape,
  black holes and gravity as the weakest force},''
  \href{http://dx.doi.org/10.1088/1126-6708/2007/06/060}{{\em JHEP} {\bfseries
  06} (2007) 060}, \href{http://arxiv.org/abs/hep-th/0601001}{{\ttfamily
  arXiv:hep-th/0601001}}.

\bibitem{Adams:2006sv}
A.~Adams, N.~Arkani-Hamed, S.~Dubovsky, A.~Nicolis, and R.~Rattazzi,
  ``{Causality, analyticity and an IR obstruction to UV completion},''
  \href{http://dx.doi.org/10.1088/1126-6708/2006/10/014}{{\em JHEP} {\bfseries
  10} (2006) 014}, \href{http://arxiv.org/abs/hep-th/0602178}{{\ttfamily
  arXiv:hep-th/0602178}}.

\bibitem{vanBeest:2021lhn}
M.~van Beest, J.~Calder\'on-Infante, D.~Mirfendereski, and I.~Valenzuela,
  ``{Lectures on the Swampland Program in String Compactifications},''
  \href{http://dx.doi.org/10.1016/j.physrep.2022.09.002}{{\em Phys. Rept.}
  {\bfseries 989} (2022) 1--50},
  \href{http://arxiv.org/abs/2102.01111}{{\ttfamily arXiv:2102.01111
  [hep-th]}}.

\bibitem{Lee:1977yc}
B.~W. Lee, C.~Quigg, and H.~B. Thacker, ``{The Strength of Weak Interactions at
  Very High-Energies and the Higgs Boson Mass},''
  \href{http://dx.doi.org/10.1103/PhysRevLett.38.883}{{\em Phys. Rev. Lett.}
  {\bfseries 38} (1977) 883--885}.

\bibitem{Lee:1977eg}
B.~W. Lee, C.~Quigg, and H.~B. Thacker, ``{Weak Interactions at Very
  High-Energies: The Role of the Higgs Boson Mass},''
  \href{http://dx.doi.org/10.1103/PhysRevD.16.1519}{{\em Phys. Rev. D}
  {\bfseries 16} (1977) 1519}.

\bibitem{Caron-Huot:2021rmr}
S.~Caron-Huot, D.~Mazac, L.~Rastelli, and D.~Simmons-Duffin, ``{Sharp
  boundaries for the swampland},''
  \href{http://dx.doi.org/10.1007/JHEP07(2021)110}{{\em JHEP} {\bfseries 07}
  (2021) 110}, \href{http://arxiv.org/abs/2102.08951}{{\ttfamily
  arXiv:2102.08951 [hep-th]}}.

\bibitem{Caron-Huot:2020cmc}
S.~Caron-Huot and V.~Van~Duong, ``{Extremal Effective Field Theories},''
  \href{http://dx.doi.org/10.1007/JHEP05(2021)280}{{\em JHEP} {\bfseries 05}
  (2021) 280}, \href{http://arxiv.org/abs/2011.02957}{{\ttfamily
  arXiv:2011.02957 [hep-th]}}.

\bibitem{Caron-Huot:2021enk}
S.~Caron-Huot, D.~Mazac, L.~Rastelli, and D.~Simmons-Duffin, ``{AdS bulk
  locality from sharp CFT bounds},''
  \href{http://dx.doi.org/10.1007/JHEP11(2021)164}{{\em JHEP} {\bfseries 11}
  (2021) 164}, \href{http://arxiv.org/abs/2106.10274}{{\ttfamily
  arXiv:2106.10274 [hep-th]}}.

\bibitem{Arkani-Hamed:2021ajd}
N.~Arkani-Hamed, Y.-t. Huang, J.-Y. Liu, and G.~N. Remmen, ``{Causality,
  unitarity, and the weak gravity conjecture},''
  \href{http://dx.doi.org/10.1007/JHEP03(2022)083}{{\em JHEP} {\bfseries 03}
  (2022) 083}, \href{http://arxiv.org/abs/2109.13937}{{\ttfamily
  arXiv:2109.13937 [hep-th]}}.

\bibitem{Cheung:2014ega}
C.~Cheung and G.~N. Remmen, ``{Infrared Consistency and the Weak Gravity
  Conjecture},'' \href{http://dx.doi.org/10.1007/JHEP12(2014)087}{{\em JHEP}
  {\bfseries 12} (2014) 087}, \href{http://arxiv.org/abs/1407.7865}{{\ttfamily
  arXiv:1407.7865 [hep-th]}}.

\bibitem{Arkani-Hamed:2017jhn}
N.~Arkani-Hamed, T.-C. Huang, and Y.-t. Huang, ``{Scattering amplitudes for all
  masses and spins},'' \href{http://dx.doi.org/10.1007/JHEP11(2021)070}{{\em
  JHEP} {\bfseries 11} (2021) 070},
  \href{http://arxiv.org/abs/1709.04891}{{\ttfamily arXiv:1709.04891
  [hep-th]}}.

\bibitem{Maldacena:2022ckr}
J.~Maldacena and G.~N. Remmen, ``{Accumulation-point amplitudes in string
  theory},'' \href{http://dx.doi.org/10.1007/JHEP08(2022)152}{{\em JHEP}
  {\bfseries 08} (2022) 152}, \href{http://arxiv.org/abs/2207.06426}{{\ttfamily
  arXiv:2207.06426 [hep-th]}}.

\bibitem{Haber:2019sgz}
H.~E. Haber, ``{Useful relations among the generators in the defining and
  adjoint representations of SU(N)},''
  \href{http://dx.doi.org/10.21468/SciPostPhysLectNotes.21}{{\em SciPost Phys.
  Lect. Notes} {\bfseries 21} (2021) 1},
  \href{http://arxiv.org/abs/1912.13302}{{\ttfamily arXiv:1912.13302
  [math-ph]}}.

\bibitem{Pasukonis:2005db}
J.~Pasukonis, ``{Gravitational scattering of massless scalars in QFT and
  superstring theory},'' \href{http://dx.doi.org/10.1002/prop.200510249}{{\em
  Fortsch. Phys.} {\bfseries 53} (2005) 1011--1029},
  \href{http://arxiv.org/abs/hep-th/0506065}{{\ttfamily arXiv:hep-th/0506065}}.

\end{thebibliography}\endgroup
\nocite{*}
}

\end{document}